# Experimental Investigation of the Hyperfine Structure of Tm I with Fourier Transform Spectroscopy, Part A: in the visible wavelength range (400 nm – 700 nm)

Şeyma Parlatan[1,2], İpek Kanat Öztürk[3], Gönül Başar[3], Günay Başar[4], Ruvin Ferber[5], Sophie Kröger[6]

[1]*Graduate School of Engineering and Sciences, Istanbul University, TR-34452 Beyazıt, Istanbul, Turkey*
[2]*Istinye University, Vocational School of Health Services,TR-34010,Zeytinburnu, Istanbul, Turkey*
[3]*Istanbul University, Faculty of Science, Physics Department, TR-34134 Vezneciler, Istanbul, Turkey*
[4]*Istanbul Technical University, Faculty of Science and Letters, Physics Engineering Department, TR-34469 Maslak, Istanbul, Turkey*
[5]*Laser Centre, University of Latvia, Jelgavas Street 3, LV-1004 Riga, Latvia*
[6]*Hochschule für Technik und Wirtschaft Berlin, Fachbereich 1, Wilhelminenhofstr. 75A, Berlin D-12459, Germany*

**Abstract**

The spectra emitted by a thulium hollow-cathode discharge lamp with argon or neon as inert gases have been recorded with a Bruker IFS 125 HR Fourier Transform spectrometer in the visible wavelength region from 400 nm to 700 nm. The paper presents the investigation of the hyperfine-structure splitting of 50 spectral lines of neutral thulium. As a result, the magnetic dipole hyperfine structure constant *A* for 20 fine structure levels of Tm I have been determined for the first time.

1. ## Introduction

Thulium (Tm), with atomic number 69, is the thirteenth element in the series of lanthanides. Its ground-state electronic configuration is $4f^{13}\,6s^2$. Due to the partially filled *f*-electron shells, the lanthanides have very complex fine structure spectra. Tm belongs to the few elements with only one stable isotope. The nuclear spin of this stable isotope, $^{169}$Tm, is $I = 1/2$. The consequence of this nonzero but small nuclear spin is that Tm lines have a very simple hyperfine structure (hfs): each fine-structure transition splits into only three or four hfs transitions.

The atomic structure of neutral Tm has been studied both experimentally and theoretically by several authors since the 1950s using different spectroscopic methods. The following paragraph gives an overview of these studies. The spectrum of Tm has been examined by Lindenberger [1], Bordarier et al. [2], Kuhl [3] and Brandt and Camus [4] using Fabry-Perot interferometers. Ritter [5] and Giglberger et al. [6] investigated the nuclear magnetic dipole moment of Tm with the atomic-beam magnetic resonance technique. Sugar et al. [7] analysed the spectrum of neutral Tm over the whole range from the ultraviolet (UV) to the near infrared (NIR) regions and re-investigated the fine-structure energy levels. Martin et al. [8] summarized all previous investigations. Different laser spectroscopic methods for hfs measurements have been used by Van Leeuwen et al. [9], Angelova et al. [10], Childs et al.

[11] Pfeufer [12–14], Krüger [15], Kronfeldt et al. [16], Kröger [17], Kröger et al. [18], Zhan-Kui et al. [19], Zhou et al. [20], Başar et al [21], and Fedorov et al. [22]. Tripplet et al. [23] investigated the Mössbauer effect of Tm. Wickliffe et al. [24] determined atomic transition probabilities for Tm I and Tm II using a Fourier Transform (FT) spectrometer. Tian et al. [25] determined the lifetimes of several levels for Tm I and Tm II. Nakamura et al. [26] studied the Zeeman effect of Tm I. Theoretical or semi-empirical investigations of the fine structure and hfs of Tm I have been performed by Bordarier et al. [2], Camus [27], Brandt and Camus [4], Kröger [17], and Biemont et al. [28].

Many of these studies contributed to examining the hfs of Tm I. However, there are still many hfs constants waiting to be measured. In the present study, we determined experimentally for the first time the hfs constant $A$ of 20 energy levels by investigating spectral lines in the visible wavelength range.

This work is supplemented by an investigation of Tm FT spectra in the near infrared wavelength range [29].

## 2. Experiment details

FT emission spectra of the element Tm were investigated in the visible wavelength range. The Tm plasma has been produced in a hollow-cathode discharge lamp, which has been cooled with liquid nitrogen. For more details concerning the hollow cathode, see [30]. In the hollow cathode tube, first high vacuum has been created and then Ar inert gas has been backfilled to a pressure of a few millibar. The tube has been operated at a discharge current of about 70 mA to provide a plasma environment. After recording the Tm-Ar spectrum, the experiment has been repeated once more using Ne as an inert gas. Thus, we have been able to distinguish the Tm lines from the inert gas lines. The radiation emitted from the hollow-cathode discharge has been directed to the FT Spectrometer with mirrors and lenses. The measurements have been performed at the Laser Centre of the University of Latvia by a Bruker IFS 125HR FT spectrometer operated with an entrance aperture of about 1 mm and a spectral resolution of 0.025 cm$^{-1}$. The radiation in the visible wavelength range has been detected by a Hamamatsu R928 photomultiplier tube with a wide spectral sensitivity range. Each spectrum has been obtained by averaging typically 100 scans. The wave number calibration of the spectra has been checked and confirmed by using Ar or Ne lines, respectively, present in the hollow-cathode discharge.

## 3. Hyperfine Structure Analysis

Since the nuclear spin $I$ of $^{169}$Tm is equal to 1/2, the nucleus has no electric quadrupole moment and only the magnetic dipole interaction leads to a hfs splitting. Every fine structure level is split into just two hfs levels. While lines with $\Delta J = \pm 1$ are split into three hfs components, lines with $\Delta J = 0$ are split into four hfs components; $J$ describes the total angular momentum of all electrons in the atom. The two main components with $\Delta J = \Delta F$ have high intensity whereas the one or two components with $\Delta J = \Delta F \pm 1$ have low intensity; $F$ describes the total angular momentum of the atom.

Due to the relatively small nuclear magnetic-dipole moment of the isotope $^{169}$Tm, $\mu_I = -0.2316\,\mu_N$ [31], the hfs splitting for most lines is small compared to the line width in our experimental setup. The line width results from the Doppler broadening in the hollow cathode lamp and from the apparatus function of the FT spectrometer. It varies between 1200 and 1600 MHz in the visible wavelength range. As a consequence, only a few Tm lines show clearly separated main peaks (with $\Delta J = \Delta F$) in our spectra.

As mentioned above, the aim of this study is to determine heretofore unknown magnetic dipole hfs constants $A$ for selected levels. For this purpose, we defined three criteria according to which the spectral lines have been selected for investigation in the current study:

- Only spectral lines with two clearly separated main peaks have been considered.
- The magnetic-dipole hfs constant $A$ of one of the two energy levels of the transition has to be known from the literature.
- The set of spectral lines has been chosen so that the unknown hfs constant $A$ for a particular energy level could be obtained from at least two lines.

We chose 51 transitions in the wavelength range between 400 and 700 nm (one of them has been excluded later, as explained below). They are listed in Table 1 together with their classifications. The classification of the lines has been accomplished with the classification program *Element* [32, 33] based on the fine-structure level energies of atomic and singly ionized Tm together with the parity, the quantum number *J*, and the hfs data, if available.

In Table 1, the level energies are given according to [8]. The given wavenumbers $\sigma$ represent the centre of gravity of the hfs, that results from the fit for our FT spectrum (as discussed below). The corresponding air wavelengths are calculated from the wavenumber values according to [34]. In columns two and three, the intensity according to [7] and the signal-to-noise ratio (SNR) in our Tm-Ar spectrum are given, respectively. Lines whose intensities are not given in the second column are new lines that are not found in the literature.

Examples of the line structures are shown in Figure 1. For the transition with $\Delta J = 0$ in Figure 1a, two weak ($\Delta F \pm 1$) hfs components with the same intensity and two strong componenets ($\Delta F = 0$) occur. Generally, the two weak hfs components are either to the left and right of the two strong components (as in Figure 1a) or both lie between the two strong components. In Figure 1b a line structure for $\Delta J = \pm 1$ with two strong and one weak hfs components is presented.

In the 51 selected transitions, 20 different levels with unknown $A$ constant values are involved. In Table 1, for each transition the level under examination is printed in bold. In almost all cases this is the upper level. Only for transitions to the level 13119.610 cm$^{-1}$ it is the lower level.

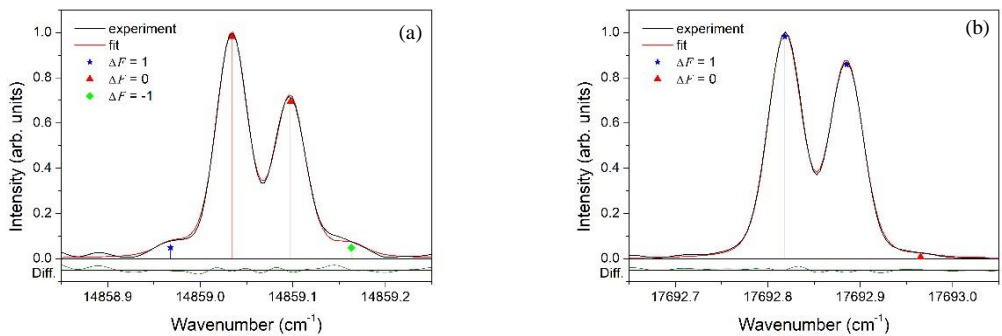

**Figure 1:** Experimental spectrum and the corresponding best fit. The residual Diff is given below the line profile.
**a)** transition 39277.087 cm$^{-1}$ ($J = 5/2$) → 24418.018 cm$^{-1}$ ($J = 5/2$) at $\sigma = $ 14859.06 cm$^{-1}$.
**b)** transition 37159.51 cm$^{-1}$ ($J = 15/2$) → 19466.663 cm$^{-1}$ ($J = 13/2$) at $\sigma = $ 17692.85 cm$^{-1}$.

**Table 1:** Tm I lines measured by means of FT spectroscopy that have been analysed in order to determine the hfs constants.

| line | | | | upper level | | | lower level | | |
|---|---|---|---|---|---|---|---|---|---|
| $\lambda_{air}$ /nm | int. | SNR | $\sigma$ / cm$^{-1}$ | $E_{up}$/cm$^{-1}$ | $P_{up}$ | $J_{up}$ | $E_{low}$/cm$^{-1}$ | $P_{low}$ | $J_{low}$ |
| 400.0568 | 40 | 21 | 24989.38 | **42742.008** | o | 7/2 | 17752.634 | e | 5/2 |
| 414.6128 | 40 | 17 | 24112.09 | **44518.91** | e | 5/2 | 20406.840 | o | 5/2 |
| 431.0401 | 30 | 13 | 23193.17 | **42742.008** | o | 7/2 | 19548.834 | e | 5/2 |
| 439.5955 | 500 | 595 | 22741.80 | **38012.795** | o | 13/2 | 15271.002 | e | 15/2 |
| 439.8572 | 40 | 20 | 22728.27 | **39470.489** | o | 5/2 | 16742.237 | e | 7/2 |
| 441.4800 | 10 | 12 | 22644.72 | **39386.959** | o | 5/2 | 16742.237 | e | 7/2 |
| 443.0863* | 50 | 118 | 22562.63 | 35682.251 | o | 7/2 | **13119.610** | e | 9/2 |
| 444.2732 | 400 | 224 | 22502.35 | **39244.584** | o | 9/2 | 16742.237 | e | 7/2 |
| 444.3283 | 4 | 7 | 22499.56 | **44919.302** | e | 9/2 | 22419.764 | o | 9/2 |
| 447.9938 | 30 | 26 | 22315.48 | **39658.852** | o | 5/2 | 17343.374 | e | 7/2 |
| 457.5318 | 30 | 33 | 21850.28 | **39602.902** | o | 3/2 | 17752.634 | e | 5/2 |
| 460.9348 | 50 | 92 | 21688.96 | **37276.778** | o | 11/2 | 15587.811 | e | 11/2 |
| 462.1716 | 300 | 157 | 21630.93 | **39244.584** | o | 9/2 | 17613.659 | e | 9/2 |
| 464.4578 | 300 | 230 | 21524.45 | **39277.087** | o | 5/2 | 17752.634 | e | 5/2 |
| 471.9269 | 30 | 45 | 21183.80 | **44518.91** | e | 5/2 | 23335.111 | o | 7/2 |
| 481.1637 | 50 | 39 | 20777.14 | **43196.848** | e | 11/2 | 22419.764 | o | 9/2 |
| 483.5898 | 5 | 32 | 20672.91 | **44919.302** | e | 9/2 | 24246.425 | o | 7/2 |
| 486.2932 | 30 | 135 | 20557.98 | **38012.795** | o | 13/2 | 17454.818 | e | 13/2 |
| 494.8352 | 100 | 266 | 20203.11 | **37657.928** | o | 11/2 | 17454.818 | e | 13/2 |
| 498.5127 | 200 | 122 | 20054.07 | **39602.902** | o | 3/2 | 19548.834 | e | 5/2 |
| 502.2458 | 4 | 8 | 19905.02 | **39658.852** | o | 5/2 | 19753.830 | e | 7/2 |
| 506.7463 | 30 | 74 | 19728.24 | **39277.087** | o | 5/2 | 19548.834 | e | 5/2 |
| 507.3519 | 100 | 451 | 19704.69 | **37159.51** | o | 15/2 | 17454.818 | e | 13/2 |
| 509.2013 | 4 | 6 | 19633.13 | **39386.959** | o | 5/2 | 19753.830 | e | 7/2 |
| 512.0672 | 150 | 137 | 19523.25 | **39277.087** | o | 5/2 | 19753.830 | e | 7/2 |
| 512.4514 | 7 | 28 | 19508.61 | **38499.016** | o | 9/2 | 18990.406 | e | 11/2 |
| 512.7818 | 20 | 32 | 19496.04 | **39244.584** | o | 9/2 | 19748.543 | e | 9/2 |
| 523.3924 | 20 | 98 | 19100.81 | **35557.70** | o | 15/2 | 16456.913 | e | 17/2 |
| 523.4797 | 7 | 8 | 19097.62 | 32217.195 | o | 9/2 | **13119.610** | e | 9/2 |
| 526.4273 | | 6 | 18990.69 | **42325.81** | e | 7/2 | 23335.111 | o | 7/2 |
| 533.1715 | 4 | 18 | 18750.48 | **38499.016** | o | 9/2 | 19748.543 | e | 9/2 |
| 533.7162 | | 12 | 18731.34 | **41151.10** | e | 9/2 | 22419.764 | o | 9/2 |
| 541.3725 | 100 | 349 | 18466.44 | **37159.51** | o | 15/2 | 18693.074 | e | 15/2 |
| 542.1666 | 3 | 13 | 18439.39 | **37276.778** | o | 11/2 | 18837.385 | e | 9/2 |
| 542.6505 | 10 | 26 | 18422.95 | **37276.778** | o | 11/2 | 18853.823 | e | 11/2 |
| 543.7783 | | 10 | 18384.74 | **42325.81** | e | 7/2 | 23941.071 | o | 9/2 |
| 549.5618 | 30 | 75 | 18191.26 | **37657.928** | o | 11/2 | 19466.663 | e | 13/2 |
| 556.6001 | 300 | 1485 | 17961.23 | 31080.846 | o | 11/2 | **13119.610** | e | 9/2 |
| 561.3231 | 70 | 66 | 17810.11 | **37276.778** | o | 11/2 | 19466.663 | e | 13/2 |
| 565.0434 | 50 | 150 | 17692.85 | **37159.51** | o | 15/2 | 19466.663 | e | 13/2 |
| 565.5846 | 5 | 20 | 17675.91 | **43196.848** | e | 11/2 | 25520.987 | o | 11/2 |
| 591.3881 | 4 | 19 | 16904.68 | **41151.10** | e | 9/2 | 24246.425 | o | 7/2 |
| 592.7922 | 5 | 38 | 16864.64 | **35557.70** | o | 15/2 | 18693.074 | e | 15/2 |
| 604.3990 | 2 | 9 | 16540.78 | **39470.489** | o | 5/2 | 22929.717 | e | 5/2 |
| 617.5297 | 400 | 774 | 16189.07 | 29308.69 | o | 9/2 | **13119.610** | e | 9/2 |
| 653.0803 | 50 | 15 | 15307.82 | **39089.533** | o | 11/2 | 23781.698 | e | 9/2 |
| 655.9511 | | 13 | 15240.83 | **39658.852** | o | 5/2 | 24418.018 | e | 5/2 |
| 664.1592 | 30 | 23 | 15052.47 | **39470.489** | o | 5/2 | 24418.018 | e | 5/2 |
| 667.8656 | 10 | 26 | 14968.94 | **39386.959** | o | 5/2 | 24418.018 | e | 9/2 |
| 672.8046 | | 30 | 14859.06 | **39277.087** | o | 5/2 | 24418.018 | e | 5/2 |
| 678.2006 | 300 | 61 | 14740.83 | **39089.533** | o | 11/2 | 24348.692 | e | 9/2 |

**Note:** $\sigma$: wavenumber of the center of gravity resulting from the hfs fit; $\lambda_{\text{air}}$: wavelengths in air, calculated from the $\sigma$ values according to [34]; int: intensity according to [7]; SNR: signal-to-noise ratio in the Tm-Ar spectrum; $E_{\text{up}}$, $E_{\text{low}}$, $p_{\text{up}}$, $p_{\text{low}}$, $J_{\text{up}}$, and $J_{\text{low}}$: level energy, parity and total electron angular momentum of the upper and lower level, respectively.
*: Resulting *A* constants not used for calculation of mean value (see text).

For the analysis of the hfs of the selected lines, the appropriate portion of the Tm-Ar-spectrum has been cut out from the entire FT spectrum for each line. Subsequently, the computer program *Fitter* [35] has been used for an iterative least-squares fit of the experimental data using a Voigt profile function for each hfs component. The fitting parameters have been the hfs constants *A* of the upper and lower levels, the center of gravity of the line, the line profile parameters, and the peak intensity parameters. The same Voigt profile parameters have been used for each hfs component within a line.

Because the hfs of the Tm lines exhibits only two strong components while the weak components partially disappear in the noise or are located below the strong components, it has been impossible to determine the hfs constants *A* of both levels of the transition from one hfs pattern. In order to determine accurately the unknown hfs constants, the hfs constants known from the literature for one of the levels has been kept fixed during the fitting procedure. A list of all hfs constants *A*, which have been fixed during the procedure is given in Table 2 with the corresponding references. If various values have been available for a level, the most accurate value has been chosen.

Additionally, during the fitting procedure, the intensity of the weak hfs components have been coupled to the intensity value of the strongest component using the theoretical intensity ratios. This is appropriate because in FT spectra the experimental intensity ratios of the hfs components generally agree well with the theoretical intensity ratios.

**Table 2:** Experimental magnetic-dipole hyperfine-structure constants $A$ of Tm I from the literature, which have been fixed during the fit procedure.

| $E$/cm$^{-1}$ | $J$ | p | $A$ /MHz | ref. |
|---|---|---|---|---|
| 15271.002 | 15/2 | e | -345.18 (23) | [13] |
| 15587.811 | 11/2 | e | -390.46 (28) | [13] |
| 16456.913 | 17/2 | e | -308.89 (17) | [13] |
| 16742.237 | 7/2 | e | -736.6 (1.0) | [11] |
| 17343.374 | 7/2 | e | -166.24 (8) | [9] |
| 17454.818 | 13/2 | e | -365.91 (55) | [13] |
| 17613.659 | 9/2 | e | -629.25 (8) | [9] |
| 17752.634 | 5/2 | e | -235.5 (1.0) | [11] |
| 18693.074 | 15/2 | e | -323.88 (30) | [13] |
| 18837.385 | 9/2 | e | -422.4 (9) | [3]* |
| 18853.823 | 11/2 | e | -476.51 (27) | [13] |
| 18990.406 | 11/2 | e | -581.4 (1.3) | [18] |
| 19466.663 | 13/2 | e | -342.84 (28) | [13] |
| 19548.834 | 5/2 | e | -57 (2) | [4]* |
| 19748.543 | 9/2 | e | -694.8 (4) | [18] |
| 19753.830 | 7/2 | e | -536.6 (9) | [4]* |
| 20406.840 | 5/2 | o | -1139.1 (4.8) | [15] |
| 22419.764 | 9/2 | o | 40.8 (1.0) | [29] |
| 22929.717 | 5/2 | e | -574 (3) | [3]* |
| 23335.111 | 7/2 | o | -819.3 (1.1) | [21] |
| 23781.698 | 9/2 | e | -306 (18) | [2]* |
| 23941.071 | 9/2 | o | -678.7 (3) | [21] |
| 24246.425 | 7/2 | o | -80.9 (1.2) | [29] |
| 24348.692 | 9/2 | e | -371.7 (1.5) | [4]* |
| 24418.018 | 5/2 | e | -659 (2) | [4]* |
| 25520.987 | 11/2 | o | -234.9 (8) | [21] |
| 29308.69 | 9/2 | o | -842.9 (1.2) | [18] |
| 31080.846 | 11/2 | o | -737.9 (2.2) | [29] |
| 32217.195 | 9/2 | o | -967.3 (1.2) | [18] |
| 35682.251 | 7/2 | o | -152.0 (2) | [19] |

**Note:** *: value converted from 10$^{-3}$ cm$^{-1}$ scale.

## 4. Results

From the analysis of the 50 spectral lines given in Table 1 we determined the hfs constants $A$ for 20 levels, of which 6 levels have even parity and 14 levels have odd parity. The results are listed in Tables 3 and 4 for the levels of even and odd parity, respectively.

All magnetic dipole hfs constants $A$ have been determined by investigating at least two spectral lines. For the estimation of the uncertainties of the individual values $A_{\text{exp}}$, two contributions have been taken into account:

- The confidence interval $\Delta A_{\text{Fitter}}$ resulting from the least-squares fit, which is calculated by the program *Fitter*.
- The errors bar $\Delta A_{\text{Fixed}}$ of the magnetic dipole hfs constant $A$ which has been fixed during the fit (see Table 2)

The root of the squares of these two values has been used for the uncertainty:

$$\Delta A_{\exp} = \sqrt{(\Delta A_{\text{Fitter}})^2 + (\Delta A_{\text{Fixed}})^2}$$

The $A_{\exp}$ values for the same level determined from different lines mostly agree within the error limits. But in several cases we found major deviations. It is presumed that the causes for these major deviations are underestimated uncertainties of the fixed $A$ values from the literature. Since it is not clear, which of these fixed $A$ values is the cause of the deviations, we cannot comment further on this issue. These deviations have been taken into account when calculating the uncertainties for the weighted mean value (see below). Only in one case of the present analysis a cause for the deviation could be found. This exception concerns the energetically low-lying level 13119.610 cm$^{-1}$, (see Table 3). For this level, four lines have been available in the visible wavelength range, three of them with relatively high intensity, (see Table 1). Nevertheless, there is a particularly large deviation between the resulting $A$ values for the line at 443.0863 nm ($\sigma$ = 22562.63 cm$^{-1}$). The profile of this line is slightly asymmetric, therefore we assume a blend with another, unknown line as a cause for the deviation. For this line, the resulting $A$ constant has not been taken into account for the calculation of the mean value.

In the penultimate column of Tables 3 and 4 the weighted mean values $A_{\text{mean}}$ of the results $A_{\exp}$ of the individual lines are given, using the reciprocal values of the squares of the uncertainties as weights $w$:

$$A_{\text{mean}} = \frac{\sum_i (A_{\exp} \cdot w_i)}{\sum_i w_i}$$

where

$$w = \frac{1}{(\Delta A_{\exp})^2}$$

In order to take into account the different contributions to the uncertainty of the weighted mean values, the following formula has been used to calculate the error bars for the weighted mean values:

$$\Delta A_{\text{mean}} = \frac{\sqrt{\sum_i (w_i \cdot \Delta A'_{\exp})^2}}{\sum_i w_i}$$

where

$$\Delta A'_{\exp} = \sqrt{(\Delta A_{\exp,i})^2 + (A_{\text{mean}} - A_{\exp,i})^2} \quad .$$

In Tables 3 and 4, for A values with large uncertainties the value and the uncertainty are given without decimal point. In contrary, in Table 2 we have followed the presentation in the references. See also comment in Table 3.

In the thesis of one of the authors, (see [17]), the results of a semi-empirical calculation of the fine structure and the hfs of configurations of even and odd parity of atomic Tm are given. For comparison, the last column of Tables 3 and 4 lists the calculated hfs constants from this reference [17]. Among the 20 new $A$ values is the one for the low-lying energy level 13119.610 cm$^{-1}$. Although this is the lowest level of even parity, no $A$ value has been known up to now. The value of $A = -441.5\,(1.2)$ MHz, which we determined for this level, is in good agreement with the result of a semi-empirical analysis of the fine structure and hfs presented in [17], where the calculated value $A_{\text{calc}} = -435$ MHz is given. For the higher-lying levels, sometimes the theoretical values show significantly larger deviations from the experimental values. This is a sign that the calculated eigenfunctions resulting from the fine structure calculation do not yet fit well and still need to be improved.

**Table 3:** New magnetic dipole hyperfine structure constants, $A$ for levels of even parity of Tm I.

| $E$ / cm$^{-1}$ | $J$ | $\sigma$ /cm$^{-1}$ | $A_{\text{exp}}$ / MHz | $A_{\text{mean}}$/ MHz | $A_{\text{calc}}$/ MHz [17] |
|---|---|---|---|---|---|
| 13119.610 | 9/2 | 16189.07<br>17961.23<br>19097.62<br>22562.63 | -441.9 (1.4)<br>-441.1 (0.6)<br>-439 (6)<br>-451.0 (1.3) * | -441.5 (1.2) | -435 |
| 41151.10 | 9/2 | 16904.68<br>18731.34 | -377 (5)<br>-383 (6) | -379 (4) | -686 |
| 42325.81 | 7/2 | 18384.74<br>18990.69 | -398 (14)<br>-393 (18) | -396 (11) | -471 |
| 43196.848 | 11/2 | 17675.91<br>20777.14 | -502 (4)<br>-499 (5) | -501 (3) | |
| 44518.91 | 5/2 | 21183.80<br>24112.09 | -339.8 (2.7)<br>-323 (8) | -338 (3) | |
| 44919.302 | 9/2 | 20672.91<br>22499.56 | -347 (7)<br>-353 (14) | -348 (7) | |

\*: $A_{\text{exp}}$ excluded from the calculation of the mean values because of blend.

**Table 4:** New magnetic dipole hyperfine structure constants $A$ for levels of odd parity of Tm I.

| $E$ / cm$^{-1}$ | $J$ | $\sigma$ /cm$^{-1}$ | $A_{\mathrm{exp}}$ / MHz | $A_{\mathrm{mean}}$/ MHz | $A_{\mathrm{calc}}$/ MHz [17] |
|---|---|---|---|---|---|
| 35557.70 | 15/2 | 16864.64 | -599.9 (1.6) | -594.7 (1.5) | -424 |
|  |  | 19100.81 | -593.6 (0.7) |  |  |
| 37159.51 | 15/2 | 17692.85 | -550.4 (0.6) | -550.6 (0.7) | -502 |
|  |  | 19704.69 | -549.6 (0.7) |  |  |
|  |  | 18466.44 | -553.1 (0.9) |  |  |
| 37276.778 | 11/2 | 17810.11 | -714.5 (1.3) | -714.0 (0.9) | -501 |
|  |  | 18422.95 | -712.3 (2.1) |  |  |
|  |  | 21688.96 | -713.8 (1.2) |  |  |
|  |  | 18439.39 | -718 (4) |  |  |
| 37657.928 | 11/2 | 18191.26 | -94.9 (1.9) | -96.1 (0.8) | -434 |
|  |  | 20203.11 | -96.3 (0.8) |  |  |
| 38012.795 | 13/2 | 20557.98 | -140.7 (0.7) | -142.4 (1.0) | -382 |
|  |  | 22741.80 | -143.5 (0.6) |  |  |
| 38499.016 | 9/2 | 18750.48 | -264 (4) | -256 (5) | -379 |
|  |  | 19508.61 | -251.7 (2.8) |  |  |
| 39089.533 | 11/2 | 14740.83 | -700.7 (1.8) | -700.6 (1.8) | -517 |
|  |  | 15307.82 | -688 (18) |  |  |
| 39244.584 | 9/2 | 19496.04 | -138.5 (2.5) | -140.1 (1.0) | -258 |
|  |  | 21630.93 | -140.5 (1.6) |  |  |
|  |  | 22502.35 | -140.2 (1.3) |  |  |
| 39277.087 | 5/2 | 14859.06 | -1291 (4) | -1280.1 (2.0) | -1110 |
|  |  | 19523.25 | -1282.0 (2.5) |  |  |
|  |  | 19728.24 | -1273 (3) |  |  |
|  |  | 21524.45 | -1278.9 (1.7) |  |  |
| 39386.959 | 5/2 | 14968.94 | -20 (14) | -17 (12) | -684 |
|  |  | 19633.13 | 4 (20) |  |  |
|  |  | 22644.72 | -25 (6) |  |  |
| 39470.489 | 5/2 | 15052.47 | 174 (6) | 175 (5) | 161 |
|  |  | 16540.78 | 175 (19) |  |  |
|  |  | 22728.27 | 179 (8) |  |  |
| 39602.902 | 3/2 | 20054.07 | -1810 (11) | -1811 (9) | -1361 |
|  |  | 21850.28 | -1813 (14) |  |  |
| 39658.852 | 5/2 | 15240.83 | 378 (11) | 389 (7) | 210 |
|  |  | 19905.02 | 386 (19) |  |  |
|  |  | 22315.48 | 394 (7) |  |  |
| 42742.008 | 7/2 | 24989.38 | -696 (5) | -696 (5) | -515 |
|  |  | 23193.17 | -697 (10) |  |  |

## 5. Conclusion and Outlook

The hfs of 50 selected lines in the high-resolution FT spectra of a Tm-Ar plasma have been investigated. All selected lines show two clearly separated strong hfs components. On this experimental basis, the magnetic dipole hfs constant *A* for 20 fine structure levels could be determined, which were not known before. The determination has been difficult due to the fact that for the isotope $^{169}$Tm the hfs only consists of two strong peaks. For some levels, the results of the individual lines do not agree with each other within the error limits. These facts are reflected in relatively large uncertainties for some *A* constants. Nevertheless, a reliable determination has been possible because the *A* constants have been determined from at least two lines per level.

The comparison with theoretical values from the literature [17] shows considerable deviations in some cases and emphasizes the importance of experimental data. Our newly determined experimental values provide a broader basis for repeating the semi-empirical analysis of fine and hyperfine structure for atomic Tm in the future.

Further lines in the infrared spectral range have been examined by a part of the team. They are published in Part B of this work [29].

**Acknowledgements**

We are grateful to Maris Tamanis and Artis Kruzins for assistance with the experiments. This study has been funded by Scientific Research Projects Coordination Unit of Istanbul University Project No.: 30048, as well as by Latvian Council of Science, project No. lzp-2020/1-088: "Advanced spectroscopic methods and tools for the study of evolved stars".

**References**

[1]   K. H. Lindenberger, „Magnetisches Kerndipolmoment und Kerndrehimpulsquantenzahl des69Tm169", *Z. Physik*, 141, 476, 1955, doi: 10.1007/BF01331891.

[2]   Y. Bordarier, R. Vetter und J. Blaise, „Étude des structures hyperfines des raies d'arc de 169Tm", *J. Phys. France*, 24, 1107, 1963, doi: 10.1051/jphys:01963002401210700.

[3]   J. Kuhl, „Hyperfeinstrukturuntersuchungen mit einem sphärischen Fabry-Perot-Interferometer mit internem Absorptionsatomstrahl im Tm I- und Eu I-Spektrum", *Z. Physik*, 242, 66, 1971, doi: 10.1007/BF01395380.

[4] H.-W. Brandt und P. Camus, „Recent hyperfine structure investigations in the configurations $4f^{13}6s^2$, $4f^{13}6s6p$, and $4f^{12}5d6s^2$ of Tm I", *Z. Physik*, 283, 309, 1977, doi: 10.1007/BF01409507.

[5] G. J. Ritter, „Ground-State Hyperfine Structure and Nuclear Magnetic Moment of Thulium-169", *Phys. Rev.*, 128, 2238, 1962, doi: 10.1103/PhysRev.128.2238.

[6] D. Giglberger und S. Penselin, „Ground-State hyperfine structure and nuclear magnetic moment of Thulium-169", *Z. Physik*, 199, 244, 1967, doi: 10.1007/BF01326434.

[7] J. Sugar, W. F. Meggers und P. Camus, „Spectrum and Energy Levels of Neutral Thulium" (eng), *Journal of research of the National Bureau of Standards. Section A, Physics and chemistry*, 77A, 1, 1973, doi: 10.6028/jres.077A.001.

[8] W. C. Martin, R. Zalubas und L. Hagan, „Atomic energy levels - the rare-earth elements", Gaithersburg, MD, 1978.

[9] K. van Leeuwen, E. R. Eliel und W. Hogervorst, „High resolution measurements of the hyperfine structure in 10 levels of Tm I", *Physics Letters A*, 78, 54, 1980, doi: 10.1016/0375-9601(80)90805-1.

[10] E. Vidolova-Angelova, G. I. Bekov, L. N. Ivanov, V. Fedoseev und A. A. Atakhadjaev, „Laser spectroscopy investigation of highly excited states of the Tm atom", *J. Phys. B: At. Mol. Phys.*, 17, 953, 1984, doi: 10.1088/0022-3700/17/6/010.

[11] W. J. Childs, H. Crosswhite, L. S. Goodman und V. Pfeufer, „Hyperfine structure of $4f^N6s^2$ configurations in $^{159}$Tb, $^{161,163}$Dy, and $^{169}$Tm", *J. Opt. Soc. Am. B*, 1, 22, 1984, doi: 10.1364/JOSAB.1.000022.

[12] V. Pfeufer, „Hyperfine structure investigations in the configuration $4f^{12}5d6s^2$ of Tm I by laser-atomic-beam spectroscopy", *Z Physik A*, 321, 83, 1985, doi: 10.1007/BF01411949.

[13] V. Pfeufer, „Hyperfine structure in the configurations $4f^{13}6s6p$ and $4f^{12}5d6s^2$ of Tm I", *Z. Physik D Atoms, Molecules and Clusters*, 2, 141, 1986, doi: 10.1007/BF01438237.

[14] V. Pfeufer, „On the hyperfine structure in the configurations $4f^m6s^2$, $4f^m5d6s$, $4f^m6s6p$, and $4f^{m-1}5d6s^2$ of the lanthanides", *Z. Physik D Atoms, Molecules and Clusters*, 4, 351, 1987, doi: 10.1007/BF01384888.

[15] J. Krüger, „Diploma Thesis". Diploma Thesis, Institut für angewante Physik, Universität Bonn, 1986.

[16] H.-D. Kronfeldt und S. Kröger, „Hyperfine structure in the configuration $4f^{13}6s7s$ of Tm I", *Z. Physik D Atoms, Molecules and Clusters*, 34, 219, 1995, doi: 10.1007/BF01437565.

[17] S. Kröger, *Hyperfeinstruktur des Thuliums und weiterer Lanthanide (Gd, Ho, Er, Yb)*. Thesis, 1996.

[18] S. Kröger, L. Tanriver, H.-D. Kronfeldt, G. Guthöhrlein und H.-O. Behrens, „High resolution measurements of the hyperfine structure and determination of a new energy level of Tm I", *Z. Physik D Atoms, Molecules and Clusters*, 41, 181, 1997, doi: 10.1007/s004600050309.

[19] J. Zhan-kui, P. Wei-xian, G. Chuan, Y. Ying-ning und Y. Hua, „Studies on hyperfine structure and isotope shift of some excited states in Yb and Tm atoms", *Acta Phys. Sin. (Overseas Edn)*, 2, 801, 1993, doi: 10.1088/1004-423X/2/11/001.

[20] Z. Zhou, L. Zhu, C. Jing, Zhang S. und F. Lin, *Acta Optica Sinica*, 13, 673, 1993.

[21] G. Başar, G. Başar, İ. K. Öztürk, F. G. Acar und S. Kröger, „Hyperfine Structure of High Lying Levels of Tm I", *Phys. Scr.*, 71, 159, 2005, doi: 10.1238/Physica.Regular.071a00159.

[22] S. A. Fedorov et al., „Improved measurement of the hyperfine structure of the laser cooling level $4f^{12}(^3H_6)5d_{5/2}6s^2$ (J=9/2) in $^{169}_{69}$Tm", *Appl. Phys. B*, 121, 275, 2015, doi: 10.1007/s00340-015-6227-5.

[23] B. B. Triplett, N. S. Dixon, L. S. Fritz und Y. Mahmud, „Electronic relaxation observed in Mössbauer hyperfine spectra of Tm metal", *Hyperfine Interact*, 72, 97, 1992, doi: 10.1007/BF02398858.

[24] M. E. Wickliffe und J. E. Lawler, „Atomic transition probabilities for Tm I and Tm II", *J. Opt. Soc. Am. B*, 14, 737, 1997, doi: 10.1364/JOSAB.14.000737.

[25] Y. Tian, X. Wang, Q. Yu, Y. Li, Y. Gao und Z. Dai, „Radiative lifetime measurements of some Tm I and Tm II levels by time-resolved laser spectroscopy", *Mon. Not. R. Astron. Soc.*, 457, 1393, 2016, doi: 10.1093/mnras/stv3015.

[26] T. Nakamura et al., „Precise g-Factor Determination of $4f^{13}6s6p$ Excited States in Tm I", *J. Phys. Soc. Jpn.*, 66, 3778, 1997, doi: 10.1143/JPSJ.66.3778.

[27] P. Camus, „Structure Hyperfine des Niveaux Pairs $4f^{12}5d6s^2 + 4f^{13}6s6p$ de Tm I", *Le Journal de Physique*, 33, 749, 1972.

[28] E. Bimont und P. Quinet, „Recent Advances in the Study of Lanthanide Atoms and Ions", *Phys. Scr.*, T105, 38, 2003, doi: 10.1238/Physica.Topical.105a00038.

[29] T. Y. Kebapci, S. Sert und Parlatan, Şeyma, et al., „Experimental Investigation of the Hyperfine Structure of Tm I with Fourier Transform Spectroscopy Part B: in the NIR wavelength range from 700 nm to 2250 nm", *to be published*, 2022.

[30] D. Messnarz und G. H. Guthöhrlein, „Investigation of the hyperfine structure of Ta I-lines (IV)", *The European Physical Journal D*, 12, 269, 2000, doi: 10.1007/s100530070022.

[31] N. J. Stone, „Table of nuclear magnetic dipole and electric quadrupole moments", *Atomic Data and Nuclear Data Tables*, 90, 75, 2005, doi: 10.1016/j.adt.2005.04.001.

[32] L. Windholz und G. H. Guthöhrlein, „Classification of Spectral Lines by Means of their Hyperfine Structure. Application to Ta I and Ta II Levels", *Phys. Scr.*, T105, 55, 2003, doi: 10.1238/Physica.Topical.105a00055.

[33] L. Windholz, „Finding of previously unknown energy levels using Fourier-transform and laser spectroscopy", *Phys. Scr.*, 91, 114003, 2016, doi: 10.1088/0031-8949/91/11/114003.

[34] P. E. Ciddor, „Refractive index of air: new equations for the visible and near infrared" (eng), *Applied optics*, 35, 1566, 1996, doi: 10.1364/AO.35.001566.

[35] A. Zeiser, S. Kröger, L. Pooyan-Weis, L. Windholz und G. Guthöhrlein, „New version of the program FITTER for fitting atomic hyperfine structure in optical spectra including the calculation of confidence intervals", *to be published*, 2022.